\documentstyle[12pt,aaspp4,epsf]{article}

\slugcomment{Appeared as ApJ 480, 705--713 (1997)}

\begin{document}

\title{Langmuir Wave Generation Through A Neutrino Beam Instability}
\author{S. J. Hardy and D. B. Melrose}
\affil{Special Research Centre for Theoretical Astrophysics
\\School of Physics, University of Sydney, NSW 2006, AUSTRALIA.
\\S.Hardy, D.Melrose@physics.usyd.edu.au}

\begin{abstract}
A standard version of a kinetic instability for the generation
of Langmuir waves by a beam of electrons is adapted to describe the
analogous instability due to a beam of neutrinos. The interaction
between a Langmuir wave and a neutrino is treated in the one-loop
approximation to lowest order in an expansion in $1/M_W^2$ in the
standard electroweak model.

 It is shown that this kinetic instability is far too weak to
occur in a suggested application to the reheating of the plasma behind a stalled shock in a
type II supernova (SN). This theory is also used to test the validity of a previous analysis
of a reactive neutrino beam instability and various shortcomings of this theory are noted.
In particular, it is noted that relativistic plasma effects have a significant effect on the
calculated growth rates, and that any theoretical description of neutrino-plasma
interactions must be based directly on the electroweak theory. The basic scalings
discussed in this paper suggest that a more complete investigation of neutrino-plasma
processes should be undertaken to look for an efficient process capable of driving the
stalled shock of a type II SN.
\end{abstract}


\section{Introduction}

Processes involving neutrinos are important in a variety of
astrophysical contexts, usually involving highly degenerate,
relativistic plasmas. For example, neutrinos provide a
cooling mechanism for red giants in the later stages of their
evolution through the generation of neutrino--antineutrino pairs from
photons (Adams, Ruderman and Woo 1963; Braaten and Segel 1993). Of
particular interest here is the intense flux of neutrinos generated in
the first few seconds of a type II supernova (SN) explosion (Bethe 1990).
There is a difficulty in understanding how the explosion occurs. The
SN shock, which is needed to eject the stellar envelope, stalls due to
energy losses through dissociation of nuclei. Heating by neutrinos is
thought to be required to drive this shock. Wilson (1985) considered
heating through direct neutrino scattering: highly beamed neutrinos
are scattered by particles in the post shock plasma and deposit energy and
momentum into this material, thereby driving the shock expansion.

An alternative mechanism for the deposition of the neutrino energy behind the SN shock is
the neutrino beam instability (Bingham et al. 1994), where the
neutrinos emitted from the SN core drive a plasma instability which produces Langmuir
turbulence. These Langmuir waves are then collisionally damped, thereby heating the
plasma, and driving the SN shock. This instability may be understood by analogy with a
more familiar electron beam instability: neutrinos propagating through a plasma emit
longitudinal waves (Langmuir waves in an unmagnetized plasma) just
as electrons do. This implies that for any plasma instability which relies on Cerenkov
emission of Langmuir waves by electrons as the driving microscopic process, there is an
analogous neutrino driven plasma instability. The simplest plasma instability is the beam
instability, where a beam of electrons (e.g., Briggs, 1964), photons (Gedalin and Eichler
1993; Melrose 1994), or neutrinos (Bingham et al. 1994), causes Langmuir waves to grow
through induced Cerenkov emission. 

In this paper a systematic development of the neutrino beam instability of
Langmuir waves is developed based on the analogy with the electron beam and
photon beam cases. A kinetic version of the neutrino beam instability is described, and the
growth rate for this instability is calculated. The results are applied to the SN shock
problem, and it is argued, contrary to the conclusions of Bingham et al. (1994). that the
instability is ineffective and unimportant.

It is known that a neutrino propagating through a medium
acquires an induced charge (Oraevsky and
Semikoz 1987; Nieves and Pal 1994) through resonant interactions with
the electrons in the medium. It is through this induced charge that
neutrinos couple to plasma waves.
Emission by a neutrino has been considered
both for longitudinal waves (Tsytovich 1964; Oraevsky and Semikoz
1987) and for transverse waves (D'Olivo, Nieves and Pal 1990; Giunti,
Kim and Lam 1991). (We note that the emission of a transverse
wave by a neutrino is kinematically allowed only if the
refractive index of the waves is greater than unity). The interaction between a
neutrino and a Langmuir wave is treated here using the standard electroweak model in the
one-loop approximation. The derivation of the probability for Cerenkov emission by a
neutrino of a Langmuir wave to lowest order in an expansion in $1/M_W^2$, where
$M_W$ is the mass of the
$W$ boson, is summarized in Appendix A. We do not
consider the effects of extensions to the standard model, such as allowing
neutrinos to have inherent electric and magnetic moments or a nonzero
intrinsic mass.

 Bingham et al. (1994; hereinafter BDSB) calculated the growth rate of a reactive
version of the neutrino beam instability and predicted that the neutrino flux from the SN
core would be sufficient to drive the instability. The theory used by BDSB was overly
restrictive, in that it was assumed that the neutrino beam was monoenergetic and that it
propagated in a cold plasma, two assumptions which increase the calculated growth rate.
We also show that the ``effective potential" neutrino-plasmon coupling used by BDSB does
not incorporate all of the necessary electroweak physics, which may also have a dramatic
effect on the growth rate of the reactive instability. We develop a kinetic version of the
reactive instability proposed by BDSB, using methods explicitly based on the electroweak
theory and drawing on techniques used in analysis of electron and photon beam
instabilities. 

In section 2 we discuss the emission of longitudinal plasmons by neutrinos and relate this
process to the theory of plasma instabilities. In section 3 the emission and absorption
coefficients are evaluated for some specific neutrino distributions. It is shown that for a
beamed distribution the absorption coefficient is negative over a range of values, implying
that wave growth occurs for a sufficiently intense neutrino beam. In section 4, the
absorption coefficient is evaluated for parameters relevant to type II SNe and it is shown
that the kinetic instability is far too weak to operate. A comparison with the work of
BDSB is then made, identifying important concerns for an explicit electroweak calculation
of the reactive version of the neutrino beam instability. We also suggest that
collisional damping completely swamps any tendency to wave growth. Natural units ($c
=\hbar = 1$) are used throughout, except where otherwise specified.

\section{Emission and absorption of Langmuir waves by neutrinos}

The treatment here of a neutrino beam instability is closely analogous with the
earlier treatments of an electron beam instability (e.g., Melrose 1986, p.\ 53)
and a photon beam instability (Melrose 1994). The time evolution of the occupation
number of the Langmuir waves is described by a kinetic equation. In general, the kinetic
equation is written in a semi-classical form involving the probability per unit
time,  $w_M({\bf k},{\bf p})$, that a particle (electron, photon or neutrino)
emit a wave quantum in an arbitrary wave mode $M$. For present purposes, we restrict
attention to the emission of the simplest longitudinal electromagnetic quantum in a
plasma, a Langmuir wave. After a description of the mechanism through which a neutrino
in a plasma may emit a Langmuir wave, the probability of this process, which is derived in
Appendix A, is written down for massless neutrinos and is simplified.

\subsection{The effective charge on the neutrino}

Cerenkov emission by a neutrino in a plasma occurs through resonant interactions of the
type shown diagramatically in Figure 1. An electron in the plasma scatters off a neutrino
via either $W$-boson or $Z$-boson exchange and in doing so emits a wave quantum. This
process is resonant when the incoming electron momentum is equal to its outgoing
momentum. The probability of the emission of a wave quantum by a neutrino in a plasma
is then calculated by averaging the probability of this resonant process over all the
electrons in the plasma. 

Alternately, one may consider the processes shown in Figure 2, and interpret the internal
electron lines as electron propagators averaged over the electrons and positrons in the
plasma (this is the approach adopted in Appendix A). There are then two contributions
from each of the diagrams of Figure 2. The first contribution is due to the vacuum
contribution of virtual electron-positron pairs. This term may does not contribute here
because the emission of a photon by a massless neutrino in vacuo is kinematically
forbidden. The second contribution is due to the real electrons and positrons in the plasma,
and is equivalent to the average over the resonant interactions shown in Figure 1.
To lowest order in $1/M_W$ the two diagrams of figure 2 may be combined to form the
4-point interaction shown in figure 3.

Nieves and Pal (1994) have calculated the induced charge on a
neutrino in a medium. The matrix element for
the interaction shown in Figure 3 (cf.\ Appendix A) may be written in the
form of an electromagnetic interaction $J^\mu(k)A_\mu(k)$, from which the 4-current,
$J^\mu$, for the neutrino is identified. The actual value of the effective
charge on the neutrino, which follows from $J^0$, is a fraction of order
$G_F n_e/ T_e$ of the charge of the electron, where
$G_F$ is the Fermi constant, $n_e$ is the electron number density, and
$T_e$ is the electron temperature as an energy. For plasma parameters
relevant to the post shock region of a type II SN, this fraction is of
order $10^{-13}$. 


The only relevant kinematic restriction to the emission process shown in Figure 2 is
conservation of 4-momentum. Denoting the 4-momenta of the neutrinos and the photons by
$(\varepsilon,{\bf p})$ and $(\omega({\bf k}),{\bf k})$, respectively, conservation of
momentum is satisfied (for
$|{\bf k}|\ll|{\bf p}|$) provided that the phase speed of the wave is less than the group
velocity of the neutrino. The phase speed of the Langmuir wave is $\omega_L({\bf k})/|{\bf
k}|$, where $\omega_L({\bf k})$ is its frequency, given by the solution of the dispersion
relation of the plasma. A general form of the dispersion relation for
Langmuir waves in a Fermi-Dirac plasma is written down in Appendix B
below. For massless neutrinos the neutrino group velocity (to lowest order in $1/M_W$) is
the speed of light. Induced effects lead to a energy-momentum relation for massless
neutrinos given by (Nieves 1989)
$\varepsilon=|{\bf p}|+\sqrt{2}G_F n_e$, where $n_e$ is the number density of
electrons. Although this implies an effective mass  and a phase speed less than
that of light, the resulting corrections are negligible in the
applications considered here. For example, for electron densities relevant to the post shock
region of a type-II SN,
$n_e=10^{36}\,{\rm m}^{-3}$,  the induced mass is less than
$10^{-7}\,{\rm eV}$ for a neutrino with energy
$\sim1\,{\rm eV}$. Thus, we may assume the simple vacuum energy-momentum relation
for the neutrinos and the kinematic restriction on the emission of Langmuir waves is 
\begin{equation}
{\omega_L({\bf k}) \over |{\bf k}|} < v_g, \label{eq:kcon}
\end{equation}
where $v_g = 1$ for massless neutrinos.

\subsection{Probability of emission}
We turn now to the calculation of the probability of emission of a Langmuir wave by
massless neutrinos in a plasma. This expression is then simplified to
lowest order in $\hbar$ in order to calculate the growth rates for the neutrino beam
instability.

The calculation of the probability per unit time of emission of a Langmuir
wave by a massless neutrino is outlined in Appendix A (cf.\  Hardy and Melrose, 1996).
The probability may be written, viz.\ equation (\ref{eq:fprob}),
\begin{equation}
w_L({\bf k},{\bf p})  = {{G_F^2 c_V^2}\over{16 \pi \alpha}}
 {|{\bf k}|^2 \omega \over {\varepsilon
\varepsilon'}}  Z(|{\bf k}|) 2 \pi \delta( \varepsilon' + \omega
- \varepsilon) 
\left[ (2 \varepsilon - \omega)^2 - |{\bf k}|^2
\right]\left(1- {\omega^2 \over {|{\bf k}|^2 }} \right)^2 , \label{eq:prob}
\end{equation}
where $G_F$ is the Fermi constant, $\alpha$ is the fine structure constant, $ Z(|{\bf
k}|)$ is a plasmon normalization factor (defined in Appendix B), with
\begin{equation}
c_V = \left\{ \begin{array}{ll}
2 \sin^2 \theta_W + {1 \over 2} & \mbox{for $\nu_e$}, \\
2 \sin^2 \theta_W - {1 \over 2} & \mbox{for $\nu_\mu$, $\nu_\tau$},
\end{array}
\right.
\end{equation}
and where $\omega = \omega_L({\bf k})$ is understood.

A convenient simplifying assumption in the calculation of the emission and
absorption coefficients is to neglect the photon recoil. The recoil is the change
in momentum, ${\bf p}$, of the neutrino due to emission of a wave quantum, with
momentum ${\bf k}$, and the recoil is small for  $|{\bf k}| \ll  |{\bf p}|$, which
also corresponds to $\omega \ll \varepsilon$. Expanding the
argument of the $\delta$--function of equation (\ref{eq:prob}) in $|{\bf k}|/|{\bf
p}|$ and retaining only the lowest order term gives
\begin{equation}
w_L({\bf k},{\bf p}) = {{G_F^2 c_V^2}\over{2 \alpha}} {{|{\bf
k}|^2 \omega} }  Z(|{\bf k}|) \left( 1- {\omega^2 \over {|{\bf k}|^2}}
\right)^2 \delta(\omega - {\bf k}\cdot{\bf v}), \label{eq:prv}
\end{equation}
where ${\bf v} = {\bf p}/\varepsilon$. Ignoring the small induced mass of the
neutrino implies $|{\bf v}| = 1$.
Equation (\ref{eq:prv}) may also be written
\begin{equation}
w_L({\bf k},{\bf p}) = C({\bf k}) \delta(\cos\chi_0 - \cos\chi),
\end{equation}
with
\begin{equation}
C({\bf k}) =  {{G_F^2 c_V^2} \over {2 \alpha}} {{|{\bf k}|^2}} Z(|{\bf k}|) {\cos\chi_0
\sin^4\chi_0},
\end{equation}
with $\cos\chi_0 = \omega/|{\bf k}|$ and where $\chi$ is the angle between ${\bf
k}$ and ${\bf v}$. 

Note that, with the neglect of the photon recoil, the
probability per unit time of emission is independent of the initial energy of the neutrino. 

\subsection{Kinetic Equations}

In the semi-classical formalism used here, the distribution of plasmons
is described by their occupation number, $N_L({\bf k})$,  and the neutrinos
are assumed non-degenerate, with distribution function normalized according
to
\begin{equation}
n_\nu = \int {d^3\,{\bf p} \over {(2 \pi)^3}} f({\bf p}), \label{eq:nnu}
\end{equation}
where $n_\nu$ is the number density of neutrinos. The time evolution of the plasmon
distribution is governed by a kinetic equation, which includes those processes which
change the number of plasmons present at a given energy through spontaneous and
stimulated emission and by absorption. Thus, the kinetic equation may be written
\begin{equation}
{d N_L({\bf k}) \over dt} = \int d^3{\bf p} \,
w_L({\bf k},{\bf p})
 \left[ \left\{1 + N({\bf k}) \right\} f({\bf p})
- N({\bf k}) f({\bf p} - {\bf k})
\right].
\label{eq:KE}
\end{equation}
where the first term within the brackets represents spontaneous and stimulated
emission, and the second term represents the inverse process, absorption of a plasmon
(microscopic reversibility requires that the probability of emission and
absorption be equal).

 For sufficiently smooth distribution functions, $f({\bf p} - {\bf k})$ may be Taylor
expanded, and then equation (\ref{eq:KE}) gives
\begin{equation}
{d N_L({\bf k}) \over dt} = \alpha_L({\bf k}) - \gamma_L({\bf k})
N_L({\bf k}), \label{eq:kin}
\end{equation}
where
\begin{equation}
\alpha_L({\bf k}) = \int d^3 {\bf p}\, w_L({\bf p},{\bf k})\,
f({\bf
p}) \label{eq:emission}
\end{equation}
and
\begin{equation}
\gamma_L({\bf k}) = - \int d^3 {\bf p}\, w_L({\bf p},{\bf k})\, {\bf
k}\cdot{\partial f({\bf p}) \over \partial {\bf p}} \label{eq:absorption}
\end{equation}
are the emission and absorption coefficients, respectively.

These coefficients appear in explicit solutions to equation (\ref{eq:kin}). In the long time
limit, the occupation number of the plasmons goes as
\begin{equation}
N_L({\bf k}) \propto e^{-\gamma_L({\bf k}) t}.
\end{equation}
For stable distributions, such as a thermal isotropic distribution, $\gamma_L({\bf k})$ is
positive, and the plasmon distribution tends to a constant level. If
$\gamma_L({\bf k})$ is negative, wave growth may occur. However, $\gamma_L({\bf
k})<0$ is not a sufficient condition for wave growth, as there are other competing plasma
processes which remove Langmuir waves from the plasma. Two of these are Landau
damping by thermal electrons and collisional damping, with damping coefficients
denoted by
$\gamma_{\rm ld}$ and $\gamma_{\rm coll}$, respectively. For growth to occur,
$\gamma_L<0$ must satisfy
\begin{equation}
|\gamma_L| >
\gamma_{\rm coll} + \gamma_{\rm ld}. \label{eq:comp}
\end{equation}
This condition sets the threshold on the neutrino intensity required for the instability to
produce Langmuir turbulence. A particularly severe restriction is imposed by
collisional damping which is of order
$(\omega_p/\Lambda)\ln\Lambda$, where
$\Lambda$ is the number of thermal electrons per Debye sphere. For the dense
plasma around the SN core $\Lambda$ is not a particularly large number
($\Lambda\sim10^2$), so that collisional damping is a strong effect to be overcome
in order for the Langmuir waves to grow. 

We turn now to evaluating the emission and absorption coefficients for a distribution of
massless neutrinos.

\subsection{Emission coefficient}

We assume that the neutrino distribution is axisymmetric and separable
in spherical polar coordinates, $(p,\alpha,\phi)$, where
$\alpha = 0$ is the streaming direction:
\begin{equation}
f({\bf p}) = f(p) \Phi(\alpha).
\end{equation}
The plasmon momentum is written in terms of
spherical coordinates $(k, \theta, \phi')$.

To perform the integrals in equation (\ref{eq:emission}) explicitly we need
the following identity (Melrose \& Stenhouse 1977),
\begin{equation}
\int_{-1}^{1} d(\cos\alpha)\int_0^{2 \pi}d \phi\, \delta(\cos\chi_0 -
\cos\chi) = \int_{\cos\alpha_-}^{\cos\alpha_+} {2 d(\cos\alpha) \over
F(\alpha,\theta,\chi_0)}
\end{equation}
with
\begin{equation}
F(\alpha,\theta,\chi_0) = (1+ 2 \cos\alpha\cos\theta\cos\chi_0 -
\cos^2\theta - \cos^2\alpha -\cos^2\chi_0)^{1/2}, \label{eq:ide}
\end{equation}
and $\cos\alpha_\pm = \cos(\chi_0 \mp \theta)$.
Using equation (\ref{eq:ide}), we find
\begin{equation}
\alpha_L({\bf k}) = 2 C({\bf k}) g(\theta,\chi_0) \int_0^\infty p^2
f(p) dp \label{eq:ec}
\end{equation}
with
\begin{equation}
g(\theta,\chi_0) = \int_{\cos\alpha_-}^{\cos\alpha_+} d\cos\alpha
{\Phi(\alpha) \over F(\alpha,\theta,\chi_0)}. \label{eq:g}
\end{equation}

\subsection{Absorption coefficient}
To evaluate the absorption coefficient, equation (\ref{eq:absorption}), we
first write
\begin{equation}
{\bf k}\cdot{\partial \over \partial {\bf p}} = k \cos\chi {\partial \over
\partial p} + {k \over p} {(\cos\alpha \cos\chi - \cos\theta) \over
\sin\alpha} {\partial \over \partial \alpha} - {k \over p \sin^2\alpha}
\left[ {\partial \over \partial \phi} \cos \chi \right] {\partial
\over \partial \phi}. \label{eq:ddp}
\end{equation}
Our system is axisymmetric, so ${\partial / \partial \phi} = 0$.
Thus equation (\ref{eq:absorption}) gives
\begin{eqnarray}
\lefteqn{ \gamma_L({\bf k})  = - C({\bf k}) \int_0^\infty p^2 dp
\int_{-1}^1 d
\cos\alpha \nonumber } \\ & & \times \int_0^{2\pi}d\phi \,
\delta(\cos\chi_0-\cos\chi)
\left(\Delta p {\partial \over \partial p} + \Delta \theta {\partial \over
\partial \theta}\right) f(p, \alpha).
\end{eqnarray}
The quantities $\Delta p = k \cos \chi$ and $\Delta \theta =
k(\cos\alpha \cos\chi - \cos\theta)/p \sin\alpha$ are the coefficients
of the relevant derivatives in equation (\ref{eq:ddp}). Further evaluation
of the absorption coefficient involves either performing the integrals
directly or expanding the particle distribution in Legendre
polynomials (Melrose \& Stenhouse 1977). One finds
\begin{equation}
\gamma_L({\bf k}) = 2 |{\bf k}| C({\bf k}) \left(2 \cos\chi_0 -
\sin\chi_0 {\partial \over {\partial \chi_0}} \right) g(\theta,\chi_0)
 \int_0^\infty p f(p) dp. \label{eq:dc}
\end{equation}
Neutrinos of all energies interact with the waves satisfying $\chi=\chi_0$. For very
low energy neutrinos, the condition $|{\bf p}|\gg|{\bf k}|$ is not satisfied. The
contribution from these low-energy neutrinos is unimportant in the present
discussion, and the integral is extended to $|{\bf p}| = 0$.

\section{Specific neutrino distributions}

Equations (\ref{eq:ec}) and (\ref{eq:dc}) may be evaluated for any
separable axisymmetric particle distribution. Our main application of
this work is to the neutrinos emitted from the core of type II
SNe, where the neutrinos passing through a point in the post shock
region are beamed due to their distance from the SN core. However, we
first show that no isotropic distribution of massless neutrinos can cause
kinetic wave growth.

\subsection{Isotropic distributions}
An isotropic distribution of neutrinos with arbitrary distribution function,
$f(p)$ has $\Phi(\alpha) = 1$. Substituting this into equation
(\ref{eq:g}) we find that
\begin{equation}
g(\theta,\chi_0) = {\pi},
\end{equation}
and hence
\begin{equation}
\alpha_L({\bf k}) ={1 \over 2} C({\bf k}) \int_0^\infty p^2 f(p) dp
\end{equation}
and
\begin{equation}
\gamma_L({\bf k}) = C({\bf k}) |{\bf k}| \cos\chi_0  \int_0^\infty p
f(p) dp > 0.
\end{equation}
It follows that the absorption coefficient is non negative and hence
wave growth is impossible for an isotropic distribution of massless neutrinos.
This result also holds for electron and photon beam instabilities.

\subsection{Beamed distributions}
The analysis relevant to type II SNe is continued through the introduction of a particular
form of beamed neutrino distribution which allows the analytic evaluation of equations
(\ref{eq:ec}) and (\ref{eq:dc}). The absorption coefficient is then maximized (in the
negative sense) with respect to the angle of emission to determine for which direction
maximum growth may occur.

We consider a highly beamed distribution of the form
\begin{equation}
\Phi(\alpha) = {2 \over \alpha_0^2} \rm{exp}\left\{{-\alpha^2 \over 2
\alpha_0^2}\right\}, \label{eq:beam}
\end{equation}
where $\alpha_0\ll 1$ is the beam opening angle (which is the angle subtended by the
neutrinosphere on the sky for type II SNe). We need to evaluate
$g(\theta,\chi_0)$ for this distribution.
Clearly, $g(\theta,\chi_0)$ is non-zero only for $\theta - \chi_0
\approx 0$, so we expand $F(\theta,\chi_0,\alpha)$ with $\alpha \ll
1$ and $\theta - \chi_0 \ll 1$. This leads to
\begin{equation}
F(\theta,\chi_0,\alpha) \approx [\alpha^2 - (\theta - \chi_0)^2]^{1/2}
\sin\chi_0.
\end{equation}
Thus,
\begin{equation}
g(\theta,\chi_0) \approx {\sqrt{2 \pi} \over \alpha_0 \sin\chi_0} \rm{exp}
\left[{-(\theta-\chi_0)^2 \over 2 \alpha_0^2} \right], \label{eq:gb}
\end{equation}
and we also have
\begin{equation}
{\partial \over \partial \cos\chi_0} g(\theta,\chi_0) \approx {-\sqrt{2 \pi}
\over \alpha_0^3 \sin^2\chi_0} (\theta - \chi_0) \rm{exp}
\left[{-(\theta-\chi_0)^2 \over 2 \alpha_0^2} \right],
\end{equation}
where we neglect the derivative of the slowly varying
$1/\sin\chi_0$ factor.
Hence, the emission coefficient is given by equation (\ref{eq:ec}) with equation
(\ref{eq:gb}) and the absorption coefficient is given by
\begin{equation}
\gamma_L(k) = 2 C({\bf k}) |{\bf k}| \left\{ 2 \cos\chi_0 + { (\theta -
\chi_0) \over
\alpha_0^2} \right\} g(\theta,\chi_0) \int_0^\infty p
f(p) dp. \label{eq:gr}
\end{equation}
The right hand side of equation (\ref{eq:gr}) is negative and largest in magnitude
for
$\theta -\chi_0
\approx -\alpha_0$. Thus the maximum growth for Langmuir waves occurs at an
angle $\theta \approx \chi_0 - \alpha_0$ to the beam axis. The maximum growth
rate corresponds to 
\begin{equation}
\gamma_{L_{\rm max}}(k) \approx - {2 C({\bf k}) k \over \alpha_0} \sin\chi_0
\, g(\chi_0-\alpha_0,\chi_0) \int_0^\infty p f(p)dp. \label{eq:gmax}
\end{equation}
Equation (\ref{eq:gmax}) and (\ref{eq:gb}) imply $\gamma_{L_{\rm max}} \propto I_\nu / \alpha_0^2$, where $I_\nu$ is the neutrino
intensity. 

\section{Application to type II SN}
Neutrinos produced in the core of a type II SN are scattered many
times as they diffuse outwards to a region which allows
essentially free propagation (for a review of the physics of type II
SN see Bethe 1990). Thus, the neutrinos have an approximately thermal
energy spectrum and appear to be emitted from a surface enclosing the
core known as the neutrinosphere. Hence, we may write the neutrino
distribution function as
\begin{equation}
f(p) = {N \over {e^{p/T_\nu} + 1}},
\end{equation}
with
\begin{equation}
N = {120 \over {7 \pi^4}} T_\nu^{-4} I_\nu,
\end{equation}
which is defined by equation (\ref{eq:nnu}), and where $T_\nu$ and $I_\nu$ are the
temperature (in units of energy) and intensity of the neutrinos, respectively. The use of a
thermal distribution of neutrinos implies that some neutrinos in the low energy tail do not
satisfy the kinematic condition (\ref{eq:kcon}), implying that the neglect of the photon
recoil is not longer justified for these neutrinos. However, as already remarked, the
fraction of neutrinos not satisfying this condition is negligible, provided the condition
$ T_\nu \gg \hbar \omega_p$ is satisfied, as is the case in practice.

At a distance $r\gg r_\nu$ from the core, where $r_\nu$ is the radius of the
neutrinosphere, the neutrinos are confined to $\alpha\la r_\nu/r$. Hence, we
take
\begin{equation}
\alpha_0 \approx {r_\nu \over r}
\end{equation}
in equation (\ref{eq:beam}).
Evaluation of the integral in equation (\ref{eq:gmax}) gives, in ordinary units,
\begin{equation}
\gamma_{L_{\rm max}}(|{\bf k}|) \approx
{{- 5 \sqrt{2 \pi} e^{-1/2}}\over{7 \pi^3}} {{G_F^2 c_V^2} \over
\alpha} {{|{\bf k}|^2} \omega \over {\hbar c^2}} Z(|{\bf k}|) \left( 1- {\omega^2 \over {|{\bf k}|^2
c^2}}
\right)^2 { 1 \over \alpha_0^2} {I_\nu
\over T_\nu^2}. \label{eq:maxg}
\end{equation}

The Langmuir wave dispersion relation in the post shock region of a type II SN requires
careful consideration. The relevant plasma is in an awkward parameter regime. Although
degeneracy may be neglected, the plasma is neither ultrarelativistic nor nonrelativistic.
Furthermore, the kinematic condition on wave emission, $v_\phi < 1$, excludes the use of
simple, known interpolation formulae which are valid only for $v_\phi > 1$ (Braaten
and Segel, 1993). The approach adopted here is to calculate the dispersion relation,
$\omega_L({\bf k})$, and the normalization factor, $Z({\bf k})$, directly from the general
expressions for an electron-positron plasma given by Hayes and Melrose (1984). This
theory is summarised in Appendix B.

We adopt the parameters used by BDSB for the material in the
post shock region of a type II SN:
$n_e = 10^{36} \,{\rm m}^{-3}$,
$I_\nu = 3 \times 10^{33} \,{\rm W m}^{-2}$,
$r = 300 \,{\rm km}$, and $\alpha_0 = 3 \times 10^{-2}$.
We also assume a neutrino temperature of $T_\nu = 3 \,{\rm MeV}$.
The electron temperature is thought to be around $2.5\times 10^{9}\, \mbox{K}$ (Miller,
Wilson, and Mayle, 1993).

\subsection{Absorption coefficient for SN plasma}
Using the techniques outlined in Appendix B, dispersion relations for a variety of plasma
temperatures and a fixed electron density of $n_e = 10^{36} \,{\rm m}^{-3}$ are plotted in
Figure 4. This is a plot of phase speed, $v_\phi/c$, against normalized wave number, $|{\bf
k}|/2mc$. The cutoff in the dispersion relation at high $|{\bf k}|$ occurs when the intrinsic
damping of the mode (the Landau damping) becomes comparable to the frequency of the
mode. In this event the plasma does not have a normal mode at that wave momentum, as
any wave is highly damped within one wave cycle. In a nonrelativistic plasma, Landau
damping limits the range of phase velocities to $v_\phi \ge e V_e = 3 (T_e/m)^{1/2}$. As
$T_e$ increases, the range $v_{\rm min} < v_\phi < 1$ shrinks, where $v_\phi = v_{\rm
min}$ is the limit imposed by Landau damping. In the ultrarelativistic limit the allowed
range of
$v_\phi < 1$ becomes very narrow.

This has immediate consequence for the strength of the emission process of Langmuir
waves by neutrinos. Equation (\ref{eq:prob}) shows that the emission probability is
proportional to $(1-v_\phi^2)^2$ which is small for $v_\phi \approx 1$. This greatly
suppresses the absorption coefficient in these high temperature plasmas. In the
ultrarelativistic limit, the region for which $v_\phi <1$ is vanishingly small, and all
emission is forbidden.

The maximum negative absorption coefficient, equation (\ref{eq:maxg}), for $T = 2.5\times
10^{9}\,
\mbox{K}$ is plotted in figure 5 as a function of normalized wave number. It is
clear that this growth rate is far too small to produce any Langmuir turbulence over the
time scale of the 3 second neutrino burst. Indeed, this instability is too weak to overcome
even the Landau damping of the SN plasma. The growth rate calculated here is
approximately 11 orders of magnitude smaller than that calculated by BDSB, who obtained
a growth rate for a reactive instability of $\approx 10^6\,\mbox{s}^{-1}$. Part of the
difference between these results lies in the the fact that a kinetic instability has been
considered here (with a thermal distribution of neutrinos), rather than the reactive
instability considered BDSB (who assume a monoenergetic distribution of neutrinos). The
greatest part of the discrepancy lies in the supression due to the high phase speed of the
longitudinal waves of the SN plasma, which is a factor of approximately $10^6$. This does
not appear in the work of BDSB due to their artificial assumption concerning the
properties of neutrinos, as is discussed below. 

One further point regarding the collisional damping of the plasma is required. This
damping was not considered by BDSB. For the parameters of the
post schock plasma assumed by BDSB the collisional damping is
$\gamma_{\rm coll} \approx 10^{15}\,\mbox{s}^{-1}$, which is many orders of magnitude
greater than the growth rates calculated either here or in BDSB. It is not clear to the
present authors how any wave mode may be generated in a plasma with such
overwhelming collisional damping.  We conclude that the neutrino beam instability is
ineffective in the post shock region of a type II SN.

\subsection{Comparison with BDSB}

The kinetic plasma instability discussed above is different from the reactive instability
discussed by BDSB. As with most instabilities, these could be regarded as kinetic and
reactive limits of a more general instability (e.g. Briggs 1964; Melrose 1986). The
kinetic version applies when the growth rate is less than the bandwidth of the growing
waves, when the random-phase approximation applies, and the reactive version applies in
the opposite limit when a fixed-phase disturbance grows in association with localized
bunching of the particles (neutrinos).

However, this equivalence does not apply to the kinetic instability discussed here and
the reactive instability discussed by BDSB because of important differences in the
underlying treatment of the neutrinos. The theory of BDSB describes the neutrino-Langmuir
wave coupling through an effective potential in a Klein-Gordon equation for the neutrino
wave field (Bethe 1986). Thus the assumed neutrino dynamics are those of a spin-0
particle, rather than the left handed spin-$1/2$ particle of the standard model. The
consequence of this assumption is that emission of a Langmuir wave in the forward
direction is allowed in the theory of BDSB, whilst it is forbidden in the electroweak
theory due to parity considerations. This is demonstrated in equation (\ref{eq:prv}) above,
and in D'Olivo, Nieves and Pal (1996). 

We commence our comparison of the theories from equation (20) of BDSB, which may be
written in our notation as
\begin{equation}
(\Omega + i \gamma_{\rm ld})^2 - w_p^2 = -{G_F^2 w_p^4 \over 2 \pi \alpha} \int {
{\bf k}\cdot {\partial f({\bf p}) / \partial {\bf p}} \over w - {\bf k}\cdot{\bf v} + i0}
d^3{\bf p} \label{eq:BDSB1}
\end{equation}
where $\omega_p$ is the plasma frequency, $\Omega = \omega - i\gamma$ includes the
real wave frequency and the wave damping, and $i0$ denotes an infinitessimal imaginary
part (the Landau prescription). This equation describes the contribution of
neutrinos to the dispersive properties of Langmuir waves in the theory of
BDSB. Note that in the limit of no neutrino flux, equation (\ref{eq:BDSB1}) gives
$\omega = \omega_p$, which is the cold plasma dispersion relation, and $\gamma =
\gamma_{\rm ld}$, which is the Landau damping of the finite temperature plasma.

The growth rate of the reactive instability of BDSB is calculated by considering the
real part of the right hand side of equation (\ref{eq:BDSB1}). The growth rate for the
kinetic instability, in the cold plasma limit, is calculated from the imaginary part of the
right hand side of equation (\ref{eq:BDSB1}), due to the pole in the integrand. This
imaginary part may be isoloated through the Plemelj formula
\begin{equation}
{1 \over \omega - \omega_0 + i0} = {\cal P} {1\over \omega - \omega_0} - i \pi
\delta(\omega - \omega_0).
\end{equation}
This imaginary part picks out the resonant interactions given by the condition $\omega -
{\bf k}\cdot{\bf v} = 0$.

To illustrate the effect of the different assumptions made concerning the neutrino
properties, we calculate the kinetic counterpart of the reactive instability considered by
BDSB. The reactive version is obtained from equation (\ref{eq:BDSB1}) by omitting the
factor $i0$ and partially integrating before setting $f({\bf p}) \propto \delta^3({\bf p})$,
and solving the resulting algebraic equation for solutions with complex frequency. The
analogous kinetic version is obtained by retaining only the resonant part when equation
(36) is inserted in equation (35). This gives $\gamma =  \gamma_{\rm ld} -
\gamma_L^{\rm BDSB}$ where
\begin{equation}
\gamma_L^{\rm BDSB} = - {G_F^2 \omega_p^3 \over 4 \alpha} \int {\bf k}\cdot {\partial
f({\bf p}) \over \partial {\bf p}} \,\delta(w - {\bf k}\cdot{\bf v}) \,d^3{\bf p}. \label{eq:c1}
\end{equation}
The result (\ref{eq:c1}) is to be compared with equation (\ref{eq:absorption}) in the cold
plasma limit, where
$Z({\bf k}) = 1$, and $\omega = \omega_p$, which gives
\begin{equation}
\gamma_L = - {G_F^2 c_V^2 \over 4 \alpha} {\omega_p^2 |{\bf k}|^2 \over \omega}
\left(1-{\omega^2 \over |{\bf k}|^2} \right)^2
\int {\bf k}\cdot {\partial f({\bf p}) \over \partial {\bf p}} \,\delta(w - {\bf k}\cdot{\bf v})
\,d^3{\bf p}. \label{eq:c2}
\end{equation}
For $|{\bf k}| \approx \omega_p$, as assumed by BDSB, equations (\ref{eq:c1})
and (\ref{eq:c2}) differ only in the factor $(1-v_\phi^2)^2$ in
equation (\ref{eq:c2}). This is precisely the term which appears through the absence of a
right handed component ot the neutrino plane wave state, as discussed above. In
particular, in a mildly relativistic plasma where the only relevant Langmuir waves have
$v_\phi < 1$ and $1-v_\phi \ll 1$, the factor $(1-v_\phi^2)^2$ in equation (38) implies
that the growth rate for left-handed neutrinos is very much smaller than for the
boson-like neutrinos assumed by BDSB.

It is clear from the difference between equations (\ref{eq:c1}) and (\ref{eq:c2}) for $|{\bf
k}| \approx \omega_p$ that any analysis of neutrino-plasma interactions must be based on
an explicit electroweak description of the microscopic processes occuring in the plasma.
This has the added advantage of allowing the analysis to be performed at relativistic
temperatures, high densities, and for broadband neutrino distributions, which are the
major approximations in the work of BDSB. An electroweak theory description of the
reactive neutrino beam instability is in progress.

\section{Conclusion}
In principle, a sufficiently intense, highly beamed distribution of neutrinos generates
Langmuir turbulence in a manner analogous to an electron or photon beam. The kinetic
version of this instability, which is studied here, is found to be too weak an instability
to be relevant in any known terrestial or space plasma. The reactive counterpart of
this instability was suggested by BDSB in connection with the
reheating of the plasma behind the stalled shock of a type II SN. However, for simplicity,
BDSB used a boson-like model for the neutrino rather than the electroweak theory which
allows only left-handed neutrinos. From a comparison of the results of our work,
based on the electroweak theory, with that of BDSB, it is clear that the two theories
lead to quatitatively different results for Langmuir waves with phase speeds close to
the speed of light, cf. equations (\ref{eq:c1}) and (\ref{eq:c2}). As the case $v_\phi < 1$
and $1-v_\phi \ll 1$ is relevant to the SN problem, an electroweak theory of the reactive
instability needs to be constructed, allowing for finite neutrino bandwidth, a finite
temperature plasma, and correct inclusion of the neutrino spin effects. Such a theory is
presently being developed.

However,  given the strong collisional damping of the SN plasma it is
unlikely that even the reactive instability can be sufficiently strong to operate in the way
suggested by BDSB. It may be that other plasma instabilities driven by an
intense neutrino beam operate in such plasmas. A thorough investigation of the
consequences of the large induced current generated by a strong flux of neutrinos in a
dense plasma is required. Such an investigation should investigate the effect of an
intrinsic neutrino mass, and other extensions to the standard model of particle physics, to
determine whether there are any conditions under which neutrino plasma instabilities can
be of astrophysical significance.

\acknowledgments 
SJH acknowledges valuable discussions with Dr. R. Bingham and Prof. J. Dawson, and thanks
the Rutherford Appleton Laborotory for their hospitality during a recent visit. 

\appendix

\section{Calculation of the probability}

The theory of the neutrino-photon vertex, as shown in figure 3, has
been derived to first order in $1/M_W^2$ (D'Olivo, Nieves and Pal 1989; Braaten and
Segel 1993), where it is shown that the scattering matrix for longitudinal plasmon
emission may be written
\begin{equation}
M_{fi} = {{G_F} \over \sqrt{s}} { {i c_V} \over \sqrt{4 \pi \alpha}}\, \bar{\nu}
\gamma_\mu (1-\gamma_5) \nu \alpha^{\mu \nu}(k) \varepsilon_\nu(k), \label{eq:Mfi}
\end{equation}
where $\alpha^{\mu \nu}(k)$ is the polarization tensor for the
electron gas,
$\varepsilon_\nu(k)$ represents the polarization 4 vector of the
longitudinal Langmuir wave, $\alpha$ is the fine structure
constant, and 
\begin{equation}
c_V = \left\{ \begin{array}{ll}
2 \sin^2 \theta_W + {1 \over 2} & \mbox{for $\nu_e$,} \\
2 \sin^2 \theta_W - {1 \over 2} & \mbox{for $\nu_\mu$, $\nu_\tau$}. \label{eq:cv}
\end{array}
\right.
\end{equation}

As we consider coupling to longitudinal plasmons, only the
longitudinal part of the linear response tensor contributes to
equation (\ref{eq:Mfi}) (Tsytovich 1964; Braaten and Segel 1993). 
Thus we may write (Melrose 1982; Nieves and Pal 1989)
\begin{equation}
\alpha^{\mu \nu}(k) \approx \alpha^L(k) L^{\mu \nu}(k,u), \label{eq:al}
\end{equation}
where
\begin{equation}
L^{\mu \nu} = -{k^2 \over {k^2-(ku)^2}} \left\{{{k^\nu
u^\mu}\over{ku}} + {{k^\mu u^\nu}\over{ku}} - {{k^2 u^\mu u^\nu} \over
{(ku)^2}} - {{k^\mu k^\nu}\over{k^2}}\right\},
\end{equation}
and, from the dispersion relation for the plasmons (Melrose 1986),
\begin{equation}
\alpha^L(k) = -{{(ku)^2}\over{\mu_0}} = - {{\omega^2}\over \mu_0}. 
\end{equation}
The second form of the above expression is evaluated in the rest frame
of the plasma, that is, the frame in which the 4-velocity of the
plasma is $u^\mu=(1,{\bf 0})^\mu$.

We choose the temporal gauge and write the polarization 4-vector for the
plasmons as 
\begin{equation}
\varepsilon^\nu(k) = {{k^\nu - (ku) u^\nu} \over {\left[(ku)^2 -
k^2\right]^{1/2}}} = \left(0,{\bf{k} \over {|\bf{k}|}}\right)^\nu.
\label{eq:p4}
\end{equation}

Explicit evaluation of $|M_{fi}|$ through equations (\ref{eq:al}) and
(\ref{eq:p4}) gives
\begin{equation}
|M_{fi}|^2 = {G_F^2 \over 2} {c_V^2 \over {4 \pi \alpha}} |{\bf
k}|^2 \omega^2 M_{\mu \nu} N^{\mu \nu}, \label{eq:mw}
\end{equation}
with
\begin{equation}
M_{\mu \nu} = 8 \left[2 p_\mu p_\nu - k_\nu p_\mu - p_\nu k_\mu + (pk)
g_{\mu\nu} + i \varepsilon_{\alpha \nu \beta \mu} k^\alpha p^\beta\right],
\end{equation}
and
\begin{equation}
N^{\mu \nu} = \left(1,{{\omega {\bf k}}\over{|{\bf k}|^2 }}\right)^\mu \left(1,{{\omega {\bf k}}\over{|{\bf k}|^2 }}\right)^\nu,
\end{equation}
with $ \omega = \omega_L({\bf k})$ and where $p^\mu = (\varepsilon({\bf p}),{\bf p})^\mu$
and
$k^\mu = (\omega_L({\bf k}),{\bf k})^\mu$ are the 4-momenta of the neutrino and
the plasmon respectively. Conservation of 4-momentum implies
\begin{equation}
pk = {1 \over 2} k^2 \label{eq:4m}.
\end{equation}

The probability per unit time of the emission of a Langmuir wave from
a neutrino may be written
\begin{equation}
w_L({\bf p},{\bf k}) = \left| M_{fi} \right|^2 {{V |a_M({\bf
k})|^2} \over {2 \varepsilon({\bf p}) 2 \varepsilon'({\bf p})}} 2 \pi
\delta( \varepsilon'({\bf p}) + \omega_L({\bf k}) - \varepsilon({\bf
p})), \label{eq:pr}
\end{equation}
where $\varepsilon({\bf p})$ and $\omega_L({\bf k})$ denote the
neutrino and Langmuir wave dispersion relations, respectively.

Substituting equations (\ref{eq:mw}) and (\ref{eq:4m}) into equation
(\ref{eq:pr}) one has
 \begin{equation}
w_L({\bf p},{\bf k}) = {{G_F^2 c_V^2}\over{16 \pi \alpha}} {{|{\bf k}|^2 } \over \omega_p^2}
{\omega^3 \over {\varepsilon
\varepsilon'}} 2 \pi \delta( \varepsilon' + \omega
- \varepsilon)
\left[ (2 \varepsilon - \omega)^2 - |{\bf k}|^2
\right]\left(1- {\omega^2 \over {|{\bf k}|^2 }} \right)^2 \label{eq:fprob}
\end{equation}
which is the probability per unit time of the emission of a Langmuir
wave by a neutrino propagation in a medium. This expression is consistent with that
derived by D'Olivo, Nieves and Pal (1996) who calculated the spontaneous emission rate of
both longitudinal and transverse wave quanta.

\section{Longitudinal photon dispersion relations in a Fermi-Dirac plasma.}

The general dispersion relation for the longitudinal mode of an isotropic Fermi-Dirac
plasma may be expressed to first order in the fine structure constant as an integral
equation. This has been done by Hayes and Melrose (1984) and we reproduce the relevant
details of their work here. 

The frequency of a longitudinal wave at a given wavenumber is given by the solution for
$\omega = \omega_L({\bf k})$ of the dispersion equation
\begin{equation}
\omega^2 = - \mu_0 \alpha^L(k)
\end{equation}
where $\alpha^L(k)$ is the longitudinal response function. This function is given by
\begin{equation}
\alpha^L(k)  =  {e^2 \bar{n}_e \omega^2 \over m |{\bf k}|^2} +  { e^2 m \omega^2 \over 2
\pi^2 |{\bf k}|^3}  \{ {1 \over 4} (\omega^2 - |{\bf k}|^2) S^{(0)}(k) - m \omega S^{(2)}(k) +
m^2 S^{(2)}(k) \}, \label{eq:lrf}
\end{equation}
where
\begin{equation}
\bar{n}_e = 2 \int {d^3 {\bf p} \over (2\pi)^3} {m \over \varepsilon}
\bar{n}(\varepsilon)
\end{equation}
is the proper electron number density, and
\begin{equation}
\bar{n}(\varepsilon) = n^-(\varepsilon) + n^+(\varepsilon)
\end{equation}
is the sum of the electron, $n^-(\varepsilon)$, and positron, $n^+(\varepsilon)$ occupation
numbers. The plasma dispersion functions of equation (\ref{eq:lrf}) are given by
\begin{equation}
S^{(0)}(k) = \int {d\varepsilon \over m} \bar{n}(\varepsilon) \ln \Lambda_1,
\end{equation}
\begin{equation}
S^{(1)}(k) = \int {d\varepsilon \over m^2} \bar{n}(\varepsilon) \varepsilon \ln \Lambda_2,
\end{equation}
\begin{equation}
S^{(2)}(k) = \int {d\varepsilon \over m^3} \bar{n}(\varepsilon) \varepsilon^2 \ln \Lambda_1,
\end{equation}
with
\begin{equation}
\Lambda_1 = {(\varepsilon_+ - \varepsilon + \omega)(\varepsilon_+ - \varepsilon -
\omega)(\varepsilon_+ + \varepsilon - \omega)(\varepsilon_+ + \varepsilon + \omega)
\over (\varepsilon_- - \varepsilon + \omega)(\varepsilon_- - \varepsilon -
\omega)(\varepsilon_- + \varepsilon - \omega)(\varepsilon_- + \varepsilon + \omega)},
\end{equation}
\begin{equation}
\Lambda_2 = {(\varepsilon_+ - \varepsilon + \omega)(\varepsilon_+ + \varepsilon -
\omega)(\varepsilon_- - \varepsilon - \omega)(\varepsilon_- + \varepsilon + \omega)
\over (\varepsilon_- - \varepsilon + \omega)(\varepsilon_- + \varepsilon -
\omega)(\varepsilon_+ - \varepsilon - \omega)(\varepsilon_+ + \varepsilon + \omega)},
\end{equation}
and
\begin{equation}
\varepsilon_{\pm} = (\varepsilon \pm 2 |{\bf p}||{\bf k}| + |{\bf k}|^2 )^{1\over 2}.
\end{equation}

Various simplifications for the degenerate, nonrelativistic and ultrarelativistic limits
exist for equation (\ref{eq:lrf}) reducing it to either a simple equation or an algebraic
transcendental equation (Hayes and Melrose 1984; Braaten and Segel 1993). However, in
its full generality, equation (\ref{eq:lrf}) must be solved numerically for specific
electron-positron distributions. 

One further plasma quantity is required for the analysis of section 4, namely the wave
normalization factor, $Z({\bf k})$, this is given by
\begin{equation}
[Z({\bf k})]^{-1} = \left.\left[1 + {\mu_0 \over 2 \omega} {\partial \over \partial
\omega}
\alpha^L(k)\right]\right|_{\omega = \omega_L({\bf k})}
\end{equation}
and is twice the ratio of electric to total energy in the wave mode (Melrose 1986).

\begin{figure}
\centerline{\epsfbox{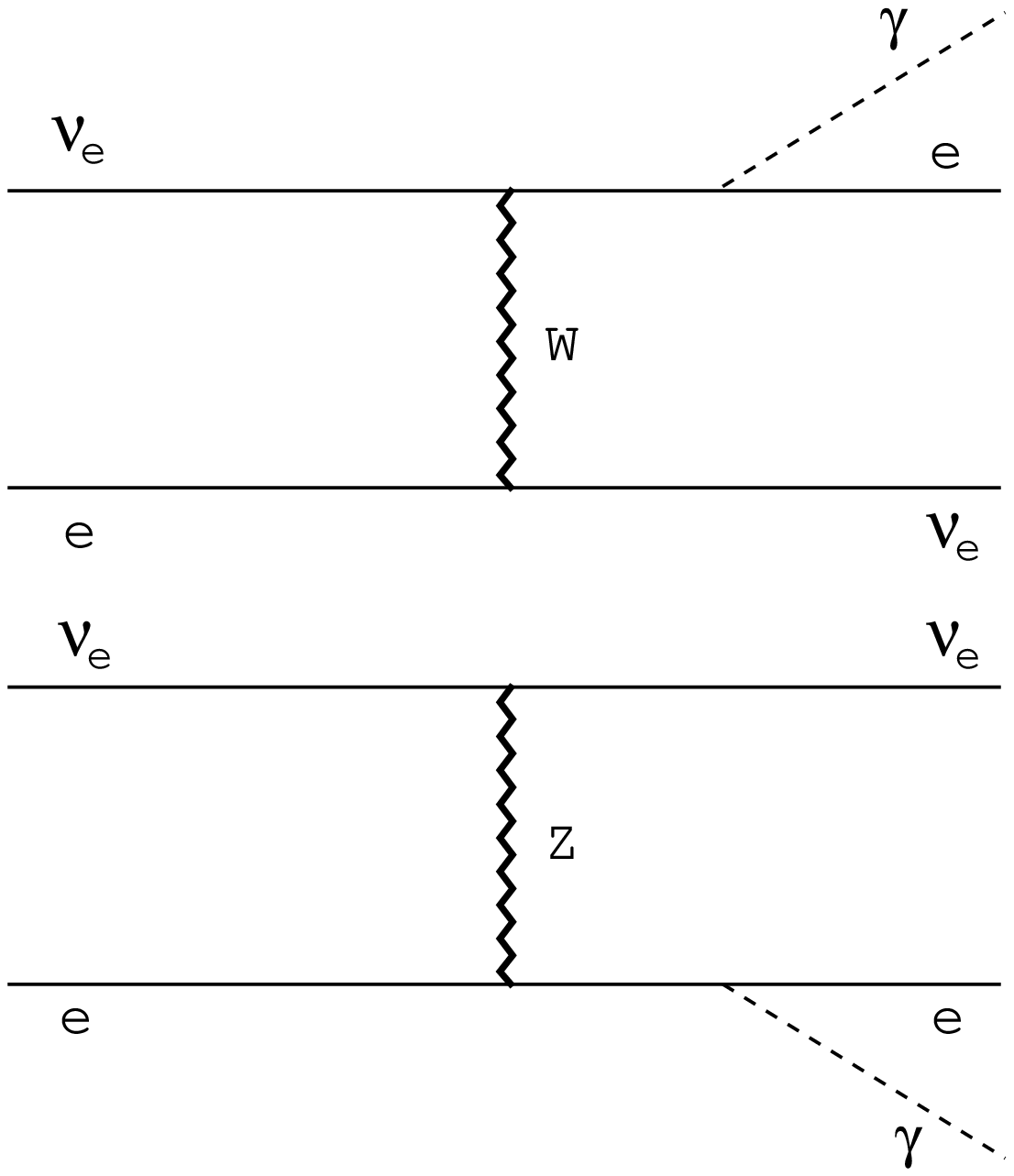}}
\caption{Feynman diagrams representing radiative decay of a neutrino
mediated by a plasma electron.}
\end{figure}

\begin{figure}
\centerline{\epsfbox{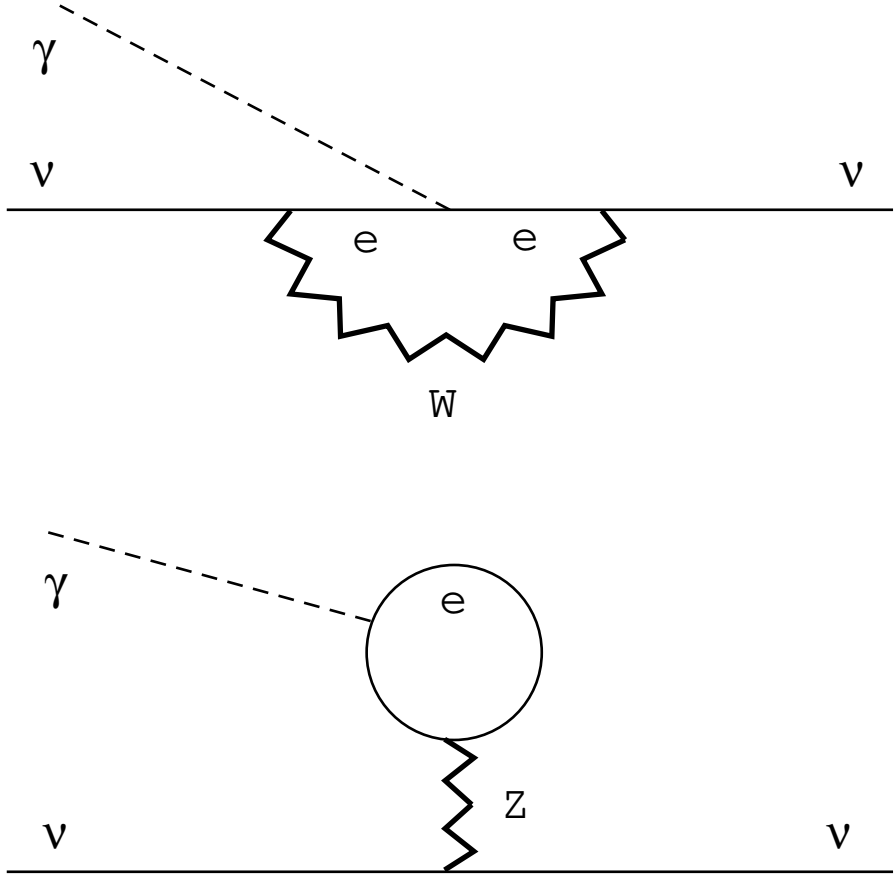}}
\caption{Feynman diagrams representing radiative decay of a neutrino
through an electron-positron pair.}
\end{figure}

\begin{figure}
\centerline{\epsfbox{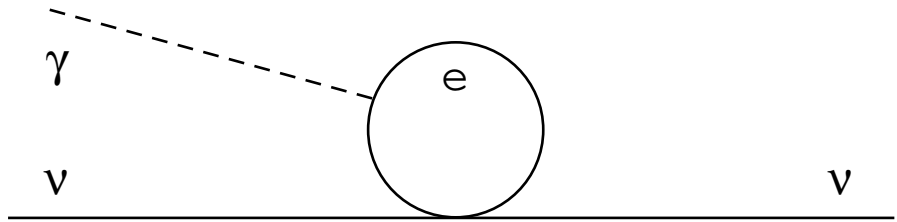}}
\caption{Feynman diagram for radiative decay of a neutrino to lowest
order in $1/M_W^2$.}
\end{figure}

\begin{figure}
\centerline{\epsfbox{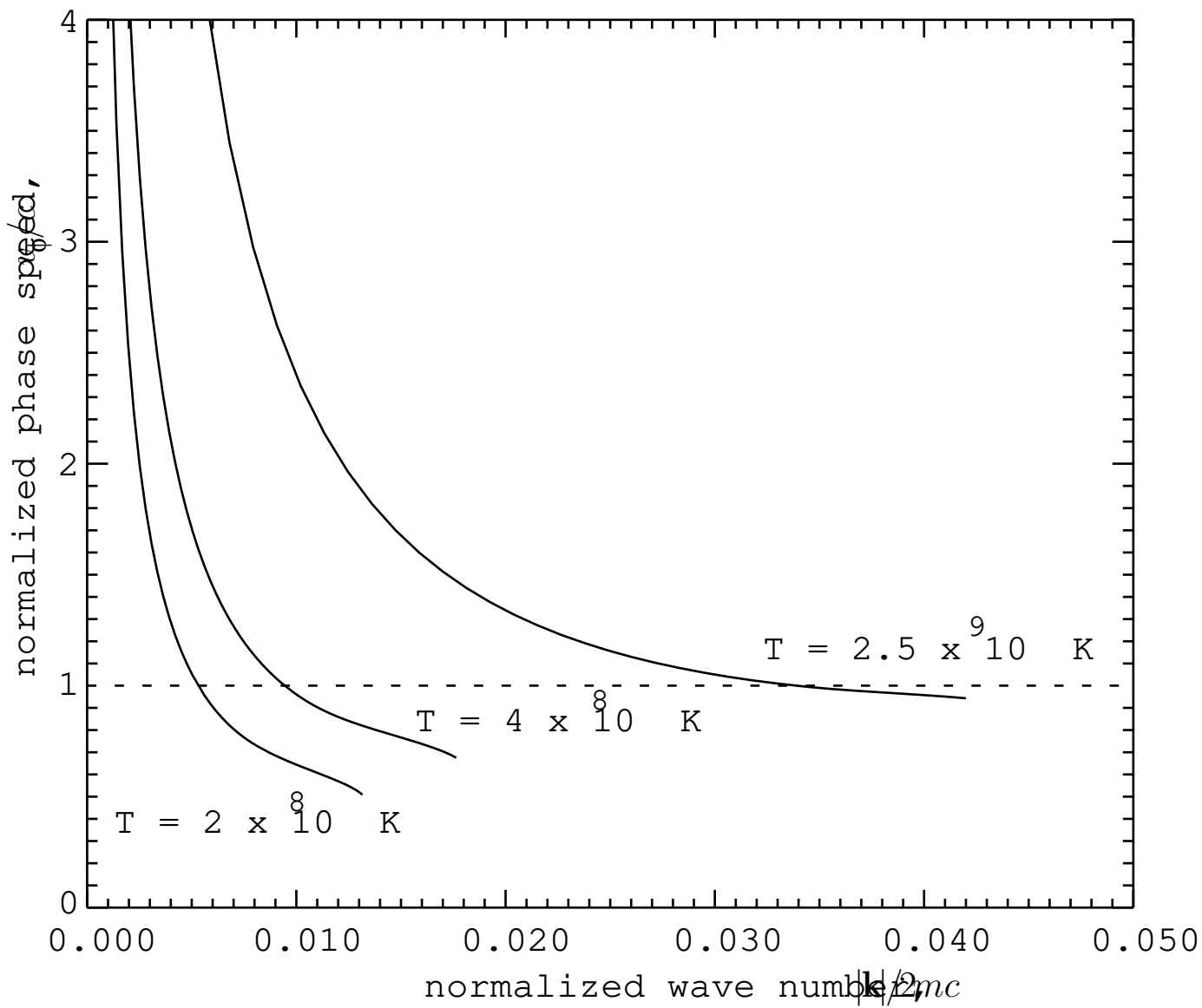}}
\caption{Plasmon dispersion relations for $n_e = 10^{36} \,{\rm m}^{-3}$. Note that
emission of Langmuir waves by neutrinos is only allowed for $v_\phi < c$. The dispersion
relation terminates at high wave number due to the destruction of the longitudinal mode
by Landau damping.}
\end{figure}

\begin{figure}
\centerline{\epsfbox{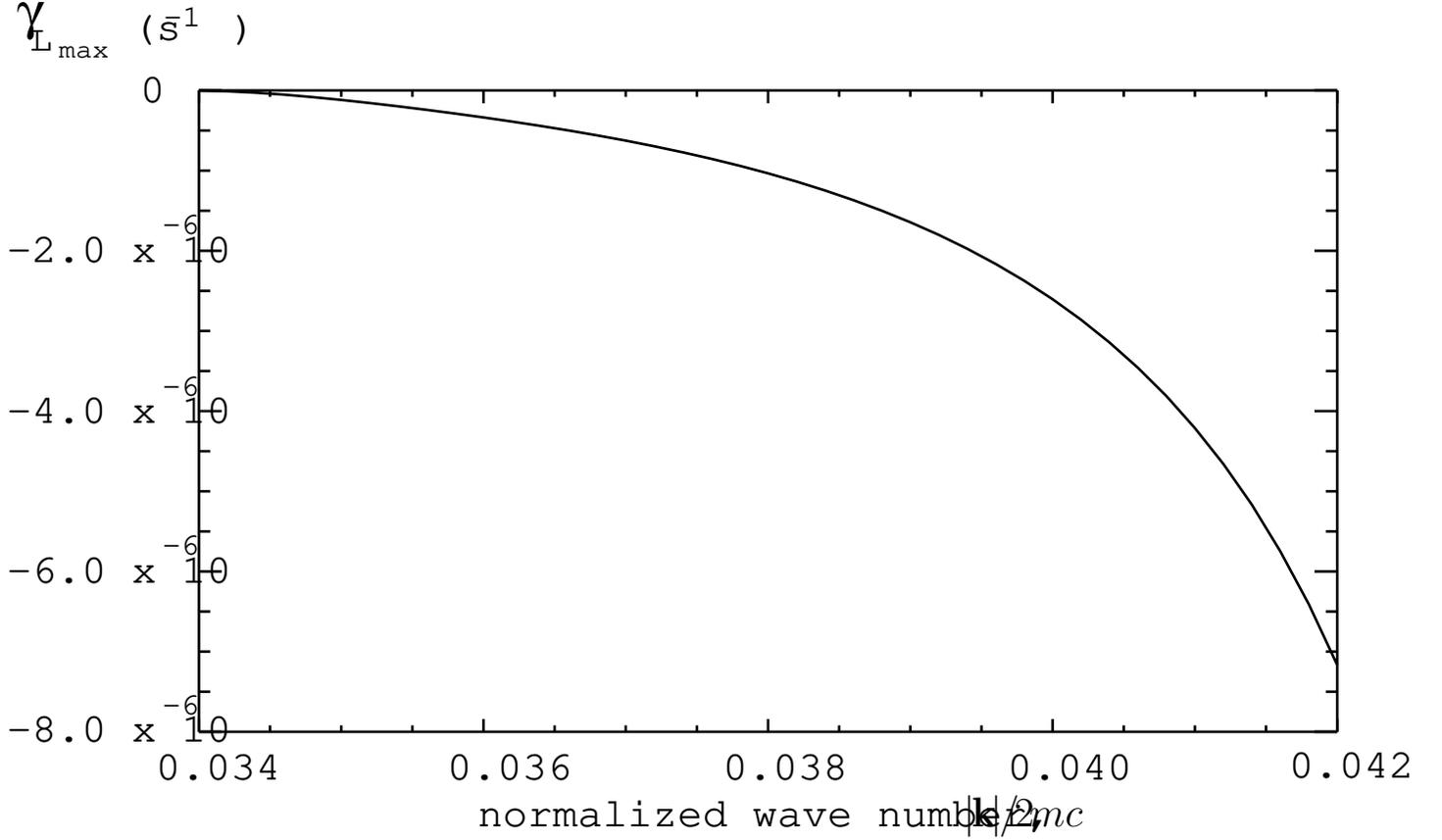}}
\caption{Maximum negative absorption for $n_e = 10^{36} \,{\rm m}^{-3}$, $I_\nu = 3
\times 10^{33} \,{\rm W m}^{-2}$, $r = 300 \,{\rm km}$, $\alpha_0 = 3 \times 10^{-2}$,
$T_\nu = 3 \,{\rm MeV}$ and $T_e = 2.5 \times10^{9}\,\mbox{K}$. The range of wave
numbers is that for which emission of a Langmuir wave is kinematically allowed, and for
which a longitudinal mode exists in the plasma (see Figure 4). The absorption coefficient
is far too small to produce Langmuir turbulence in the post shock plasma of a type II SN.}
\end{figure}

\end{document}